\newcounter{myctr}
\def\myitem{\refstepcounter{myctr}\bibfont\noindent\ifnum\themyctr>9\else\phantom{0}\fi\hangindent17pt\themyctr.\enskip}

\documentclass{ws-ijqi}
\usepackage{bbm,amsmath, amssymb,amsbsy,graphicx,subfigure,hyperref,color}

\newcommand{\gr}[1]{\boldsymbol{#1}}
\newcommand{\be}{\begin{equation}}
\newcommand{\ee}{\end{equation}}
\newcommand{\bea}{\begin{eqnarray}}
\newcommand{\eea}{\end{eqnarray}}
\newcommand{\ket}[1]{|#1\rangle}
\newcommand{\bra}[1]{\langle#1|}

\newcommand{\sig}{\gr{\sigma}}

\newcommand{\eq}[1]{Eq.~(\ref{#1})}

\begin{document}

\markboth{G. Adesso and D. Girolami}
{Gaussian Geometric Discord}

\catchline{}{}{}{}{}

\title{GAUSSIAN GEOMETRIC DISCORD}

\author{GERARDO ADESSO}

\address{School of Mathematical Sciences, University of Nottingham, University Park, Nottingham NG7 2RD, United Kingdom\\
gerardo.adesso@nottingham.ac.uk}

\author{DAVIDE GIROLAMI}

\address{School of Mathematical Sciences, University of Nottingham, University Park, Nottingham NG7 2RD, United Kingdom\\
pmxdg1@nottingham.ac.uk}

\maketitle

\begin{history}
\received{Day Month Year}
\revised{Day Month Year}
\end{history}

\begin{abstract}
We extend the geometric measure of quantum discord, introduced and computed for two-qubit states in \cite{dakic}, to quantify non-classical correlations in composite Gaussian states of continuous variable systems. We lay the formalism for the evaluation of a Gaussian geometric discord in two-mode Gaussian states, and present explicit formulas for the class of two-mode squeezed thermal states. In such a case, under physical constraints of bounded mean energy, geometric discord is shown to admit upper and lower bounds for a fixed value of the conventional (entropic) quantum discord. We finally discuss alternative geometric approaches to quantify Gaussian quadrature correlations.
\end{abstract}


\section{Introduction}

Gaussian states of quantised fields and matterlike systems constitute a tiny class in the arena of general continuous variable (CV) quantum states, yet their role is recognised as a leading one in quantum technology applications \cite{brareview}. They are defined as those states with a positive-anywhere, Gaussian-shaped Wigner distribution in phase space. As positivity of the Wigner function is sometimes regarded as a signature of ``classicality''\cite{eisertpolzik}, one might wonder how come quantum protocols can be implemented and enhanced by the use of states which do not appear to radically deviate from a classical description. However, this picture is too simplistic. A more refined notion of classicality can be adopted, whereas quantum states $\rho$ are considered as ``classical'' if their Glauber--Sudarshan $P$ representation, implicitly defined by $\rho = \int {\rm d}^2 \alpha P(\alpha) \ket{\alpha}\bra{\alpha}$ (with $\ket{\alpha}$ denoting a coherent state), is regular and nonnegative\cite{classoptP}. With respect to this criterion, a numerical study has demonstrated that essentially almost all two-mode Gaussian states are non-classical\cite{slater}. The typical example of a non-classical Gaussian state is a squeezed state.  Squeezing represents an essential ingredient to establish entanglement between two or more modes in a global Gaussian state\cite{squeezirreducible}; entanglement, in turn, constitutes an unambiguous quantum trait, ultimately enabling better-than-classical realisations of communication primitives such as quantum teleportation\cite{telep}. For this reason, the structure and distribution of bipartite and multipartite entanglement in multimode Gaussian states, and its operational significance for teleportation and related schemes,  has been extensively studied in recent years\cite{ourreview}.

One could then ask whether non-entangled Gaussian states are just ``classical''. Surprisingly enough, this question was settled only very recently. The concurrent works of Giorda and Paris\cite{Giorda_10} and Adesso and Datta\cite{AD_10} provide a clear, negative answer to that. In those studies, a paradigm more general than the entanglement-versus-separability one\cite{Werner89} has been explored to assess the quantumness of correlations in bipartite Gaussian states. This paradigm was proposed one decade ago in two pioneering independent works by Zurek and Vedral (and their collaborators)\cite{OZ,HV}. They identified signatures of quantumness stemming directly from the non-commutativity of quantum observables, which can be retrieved for instance in the fact that   local measurements on one or more subsystems of a quantum composite system typically induce a disturbance on the state of the system, at striking variance with what happens when extracting marginal information from a classical joint probability distribution. This generally happens in entangled as well as in generic separable quantum states. Precisely, this is the case for almost all quantum states, and the purely classically correlated (or ``classical'') states constitute just a null measure, nowhere dense subset of any composite Hilbert space\cite{acinferraro}. The classical states are just those with a `collapsed' wavefunction, also known as pointer states\cite{zurekreview}, whose density matrix is diagonal in some local product basis. For a more precise mathematical definition, please refer to the paper by Gharibian et al. in this special issue\cite{noiijqi}.

The evidence of quantum correlations in separable states has been mostly ignored until it was speculated that such correlations might be at the root of the quantum speed-up in computational algorithms that run over mixed qubit registers, such   as the DQC1 model\cite{dattabarbieriaustinchaves}. In an attempt to capture, operationally and/or from a more abstract perspective, the essence of general non-classical correlations, several measures have been proposed for their quantification, among which we mention Refs.~\cite{OZ,HV,piani,MID,altremisure,req,modi,dakic,genio}. For two-mode Gaussian states, detailed results are available concerning the evaluation of the quantum discord \cite{Giorda_10,AD_10} and of the (ameliorated) measurement-induced disturbance\cite{gamid}. In the case of quantum discord ${\cal D}$ \cite{OZ}, which is one of the most popular measures of non-classical correlations at the moment, a closed formula is available for general Gaussian states of two-mode CV systems, under the restriction of Gaussian local measurements performed on one of the party\cite{AD_10}, while only a semi-analytical expression exists for the quantum discord of general states of two qubits\cite{davide}. On the other hand, an intuitive geometric approach to the quantification of non-classical correlations has recently led to the introduction of the ``geometric measure of quantum discord'' (or geometric discord) $D_G$ \cite{dakic,luofu}, which is easily computable in closed form for general two-qubit states.

In this paper we propose a generalisation of the geometric discord to Gaussian states of CV systems. We define the ``Gaussian geometric discord'' (GGD) as the minimum squared Hilbert-Schmidt distance between a Gaussian state and the closest `classical-quantum' state obtained after a local generalised Gaussian positive-operator-valued-measurement  (GPOVM)  performed on one party only. Focusing on two-mode Gaussian states, we provide a covariance matrix formulation of the problem, and we solve the optimisation explicitly for the relevant subclass of two-mode squeezed thermal states: the optimal measurement in this case is shown to be  a heterodyne detection, similarly to the case of conventional quantum discord\cite{Giorda_10,AD_10}, and the GGD turns out to be symmetric under swapping of the two parties. We then compare GGD with the quantum discord for two-mode squeezed thermal states. While an upper bound on GGD can be identified exactly at fixed quantum discord, states can be found with arbitrarily small GGD for any arbitrarily large quantum discord. We recognise this as a consequence of the known result that infinitely entangled states of CV systems are dense in norm in infinite-dimensional Hilbert spaces\cite{eisertsimon}. In fact, imposing simple physical constraints of finiteness of the mean energy, we are able to recover as well a hierarchy of lower bounds on GGD versus discord. We finally discuss possible alternative generalisations of the  geometric discord\cite{dakic,luofu} to Gaussian states. If one applies, for instance, the notion of minimum Hilbert-Schmidt distance from classical states directly at the level of covariance matrices, an easily computable measure of total quadrature correlations is obtained.

 The paper is structured as follows. In section \ref{secPrelim} we recall the notion of geometric discord, and propose its generalisation to Gaussian states of CV systems. In section \ref{sec2m} we evaluate the GGD for two-mode Gaussian states, and compare it to the conventional (entropic) quantum discord. In section \ref{secQ} we discuss an alternative geometric measure of Gaussian  correlations. Finally, section \ref{secC} contains a brief summary and outlook.

\section{Geometric discord}\label{secPrelim}
The geometric measure of quantum discord $D_G$ has been recently proposed as a simple and intuitive quantifier of general non-classical correlations in bipartite quantum states\cite{dakic}. Let us consider a bipartite system $AB$ in the state $\rho_{AB}$ and imagine  to perform a local measurement $\Pi_B$ on  $B$. Almost all (entangled or separable) states $\rho_{AB}$ will be disturbed by any such measurement\cite{acinferraro}. However, for the class of so-called classical-quantum states\cite{piani,noiijqi} of the form
\begin{eqnarray}\label{cq}
\chi_{AB} = \sum _i  p_i \rho_{Ai}\otimes |i\rangle  \langle i | ,
\end{eqnarray}
where $p_i$ is a probability distribution and $\{|i\rangle\}$ is a basis for the Hilbert space of subsystem $B$, there exists at least one local measurement that leaves the state invariant.
 Denoting by $\Lambda$ be the set of classical-quantum states with respect to the $A$ versus $B$ bipartition, the geometric discord $D_G$ is defined as  the squared Hilbert-Schmidt distance between  the state $\rho_{AB}$ and the closest classical-quantum state $\chi_{AB}$ \cite{dakic},
 \begin{equation}\label{dgeom}
  D_G(\rho_{AB})= \inf_{\chi_{AB} \in \Lambda}  \|\rho_{AB} -\chi_{AB} \|_2^2\,.
  \end{equation}
where
$\|M\|_2=\sqrt{{\rm tr}(M M^\dagger)}$ stands for the $2$-norm, or Hilbert-Schmidt norm, of the matrix $M$.
The quantity $D_G (\rho_{AB})$ in \eq{dgeom}  vanishes on classical-quantum states and is upper bounded by the state purity  ${\rm tr} (\rho_{AB}^2)$ in general. Interestingly, the geometric discord can be exactly reformulated as the minimal disturbance,  measured in terms of the squared Hilbert-Schmidt distance, induced on the state $\rho_{AB}$ by any projective measurement $\Pi^B$ on subsystem $B$ \cite{luofu},
\begin{equation}\label{dgeomm}
D_G(\rho_{AB})= \inf_{\Pi_B}  \|\rho_{AB} -\Pi_B(\rho_{AB}) \|_2^2\,.
\end{equation}
 We notice that the geometric discord is in general not symmetric under  the swap of the two parties, $A \leftrightarrow B$, i.e. measuring $A$ rather than   $B$ may induce different amounts of disturbance on generic bipartite states.

In this paper we shall consider $\rho_{AB}$ to be a bipartite, $n$-mode Gaussian state $\rho_{AB}^{\cal G}$, i.e. a state with Gaussian Wigner distribution on the quantum phase space of $n$ quantised harmonic oscillators, partitioned into two blocks of $n_A$ and $n_B$ modes, respectively (with $n=n_A+n_B$). Without loss of generality, we can assume that all the first moments of the canonical phase-space operators are set to zero, because we are aiming at computing correlations between the modes, which are not affected by local displacements. The states under consideration are then entirely described by their covariance matrix $\gr\Sigma_{AB}$ of elements  $\Sigma_{ij} = {\rm tr} (\rho_{AB} \{\hat{R}_i,  \hat{R}_j\}_+)$ where ${\bf \hat{R}} = (\hat{x}_1,\hat{p}_1,\ldots,\hat{x}_n,\hat{p}_n)$ is the vector of phase-space operators satisfying the canonical commutation relations $[\hat{R}_i, \hat{R}_j] = i \Omega_{ij}$, with $\gr \Omega={{\ 0\ 1}\choose{-1\ 0}}^{\oplus n}$ being the symplectic matrix\cite{ourreview}.

We then define the Gaussian geometric discord (GGD) $D_G^{\cal G}$ of a Gaussian state $\rho_{AB}^{\cal G}$ as in \eq{dgeomm}, with $\Pi_B$ constrained to be a generalised  GPOVM $\Pi_B^{\cal G}$ , i.e. a   map  sending Gaussian states into Gaussian states. These measurements coincide with the standard toolbox of linear optics, i.e.,~can be realised using beam splitters, phase shifters, squeezers, appending ancillary vacuum modes, and performing balanced homodyne detection\cite{giedkefiurasek}.
The GGD is then, explicitly,
\begin{equation}\label{dgeommg}
D_G(\rho^{\cal G}_{AB})= \inf_{\Pi^{\cal G}_B}  \|\rho^{\cal G}_{AB} -\Pi^{\cal G}_B(\rho^{\cal G}_{AB}) \|_2^2\,,
\end{equation}
which is normalised between $0$ and $1$ for Gaussian CV states.
From now on, we  drop all the superscripts ``${\cal G}$'' and always assume, implicitly, that all the states and measurements we consider are Gaussian.

\section{Gaussian geometric discord for two-mode Gaussian states}\label{sec2m}
We shall focus on two-mode Gaussian states $\rho_{AB}$, specified by their $4 \times 4$ covariance matrix
\begin{equation}\label{sig}
\gr\Sigma_{AB} = \left(
                \begin{array}{cc}
                  \gr\alpha & \gr\gamma \\
                  \gr\gamma^T & \gr\beta \\
                \end{array}
              \right)\,,
\end{equation}
which has to be positive definite and has to comply with the uncertainty principle, $\gr\Sigma_{AB} + i \gr\Omega \ge 0$, to ensure its correspondence to a physical state.
The GPOVM on the single-mode subsystem $B$ can be written in general as $\Pi_B(\eta) = \pi^{-1} \hat{W}_B(\eta) \varpi_B \hat{W}^\dagger_B(\eta)$ where $\hat{W}_B(\eta) = \exp(\eta \hat{b}^\dagger - \eta^\ast \hat{b})$ is the Weyl operator, $\hat{b} = (\hat{x}_B + i \hat{p}_B)/\sqrt2$, $\pi^{-1}\int d^2\eta \Pi_B(\eta) = {\mathbbm{1}}$, and the seed $\varpi_B$ is the density matrix of a (generally mixed) single-mode Gaussian state with zero mean and covariance matrix $\sig_B$. After the measurement, the  state of mode $B$ is projected onto the seed $\varpi_B$, while the posterior state $\varpi_A$ of mode $A$ has a covariance matrix that is crucially not dependent on the measurement outcome\cite{giedkefiurasek}, being given by the Schur complement
\begin{equation}\label{siga}
\sig_A = \gr\alpha-\gr\gamma (\gr\beta+ \sig_B)^{-1} \gr\gamma^T\,.
 \end{equation}
 This entails that no classically correlated mixture can be realised as the outcome of a local Gaussian measurement, and the only classical-quantum two-mode Gaussian states are just uncorrelated tensor product states, as already proven in\cite{AD_10}.  The GGD for two-mode Gaussian states thus reads \begin{equation}\label{dg2tmp}
 D_G(\rho_{AB})= \inf_{\varpi_B}  \|\rho_{AB} -\varpi_A \otimes \varpi_B \|_2^2\,,\end{equation} where the single-mode states $\varpi_{A,B}$ are introduced above.

\begin{figure}[htb]
\centering
\includegraphics[width=8.5cm]{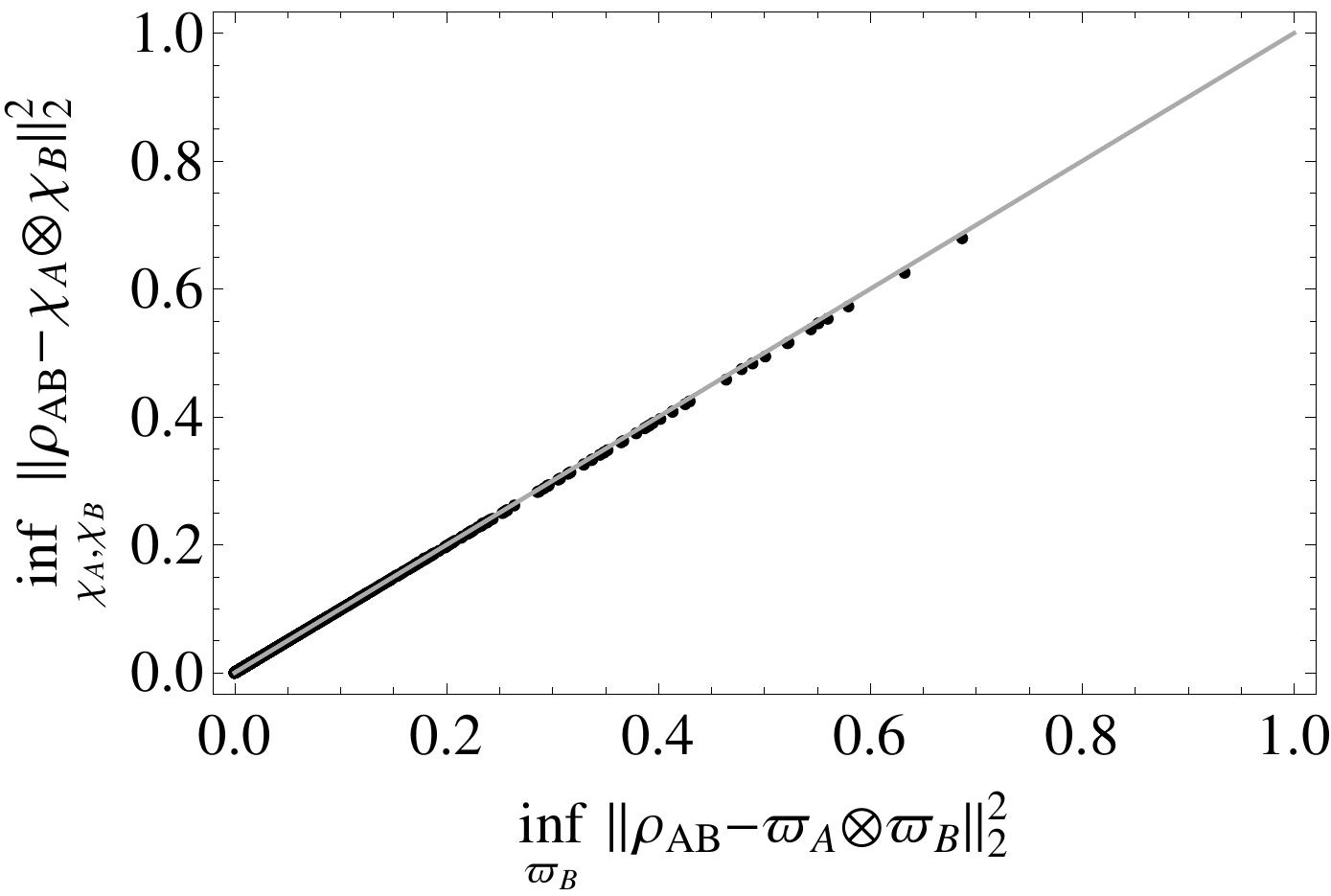}
\caption{Comparison between two possible Gaussian extensions of the geometric discord (see the main text for further details), one emerging from \eq{dgeom} (vertical axis), and the other emerging from \eq{dgeomm} (horizontal axis). The latter is what we define as GDD in \eq{dgeommg} and adopt in this paper. The points correspond to $10000$ randomly generated two-mode Gaussian states. All the quantities plotted are dimensionless.}
\label{figlinea}
\end{figure}

 We remark that in our notation, the optimisation is over $\varpi_B$ only, and there is no free parameter left in $\varpi_A$.
Let us comment on this issue in some detail. One might think to  define alternatively the Gaussian counterpart of the geometric discord as in \eq{dgeom}, by letting $\chi_{AB}$ be any possible two-mode tensor product Gaussian state, i.e. $\chi_{AB} = \chi_A \otimes \chi_B$, with completely arbitrary $\chi_A$ and $\chi_B$, and optimising over both. This class of states is  clearly  more general  than the subset of tensor product states emerging from the GPOVM analysis, which enters in our definition \eq{dg2tmp}, thus possibly enabling a further minimisation of the Hilbert-Schmidt distance compared with the optimal value of $D_G$ obtained from \eq{dg2tmp}. However, a numerical comparison [see Fig.~\ref{figlinea}] reveals that the two definitions  result in actually quite close (although not exactly coincident) values for randomly generated two-mode Gaussian states. Therefore, we  believe it more natural and  operationally wise to stick with the definition of $D_G$ that we introduced in \eq{dgeommg}. A reconcilation between Eqs. (\ref{dgeom}) and (\ref{dgeomm}) for bipartite Gaussian states would be possible only enlarging the set of operations $\Pi_B$ to general, even non-Gaussian CV measurements\cite{gamid} (and, equivalently, the set $\Lambda$ to generally non-Gaussian two-mode  classical-quantum states), which is beyond the scope of this paper.

  We now proceed with the evaluation of \eq{dg2tmp}.
  Recalling the formula for the overlap of two Gaussian states $\rho_1$ and $\rho_2$ with covariance matrices $\gr\sigma_1$ and $\gr\sigma_2$ \cite{scutaru}, ${\rm tr} (\rho_1 \rho_2) = 1/\sqrt{\det[(\sig_1+\sig_2)/2]}$, one has
\begin{equation}
\label{dg2}
D_G(\gr\Sigma_{AB})= \inf_{\sig_B}  \left\{\frac{1}{\sqrt{\det \gr\Sigma_{AB}}}+\frac{1}{\sqrt{\det (\sig_A \oplus \sig_B)}} - \frac{2}{\sqrt{\det[(\gr\Sigma_{AB}+\sig_A\oplus\sig_B)/2]}}\right\}\,,
\end{equation}
with $\gr\Sigma_{AB}$ given by \eq{sig} and $\sig_A$ given by \eq{siga}.

The measure $D_G$ is local-unitarily-invariant by definition.  On the other hand, by means of local unitary operations (symplectic transformations in phase space), the covariance matrix of every two-mode Gaussian state can be brought into a unique standard form\cite{duan00},
\begin{equation}\label{sigsf}
\gr\Sigma_{AB}=
\left(
\begin{array}{cccc}
 a & 0 & c & 0 \\
 0 & a & 0 & d \\
 c & 0 & b & 0 \\
 0 & d & 0 & b
\end{array}
\right)\,,
\end{equation}
with $a,b \ge 1$, $\sqrt{a b -1} \ge c \ge |d|$.
There is no loss of generality in restricting ourselves to states in standard form in order to proceed with the explicit computation of $D_G$. The formulas that we derive in the following are in fact valid for arbitrary two-mode Gaussian states, provided one identifies the standard form covariances in terms of the four symplectic invariants $\det\gr\alpha=a^2$, $\det\gr\beta=b^2$, $\det\gr\gamma=c d$, $\det\gr\Sigma_{AB}=(a b - c^2)(a b - d^2)$ \cite{ourreview}.

  Now, the seed state is in general a single-mode rotated, squeezed, thermal state with covariance matrix
\begin{equation}
\label{sigb}
\sig_B=\left(
\begin{array}{cc}
 m \lambda  \cos ^2(\theta )+\frac{m \sin ^2(\theta )}{\lambda } & -\frac{m
   \left(\lambda ^2-1\right) \cos (\theta ) \sin (\theta )}{\lambda } \\
 -\frac{m \left(\lambda ^2-1\right) \cos (\theta ) \sin (\theta )}{\lambda } &
   \frac{m \cos ^2(\theta )}{\lambda }+m \lambda  \sin ^2(\theta )
\end{array}
\right)\,,
\end{equation}
with $m \ge 1$ and $\lambda \ge 0$.
The task we are facing is then the minimisation of \eq{dg2} for two-mode states in standard form, over the GPOVM parameters $m$ (temperature), $\lambda$ (squeezing), and $\theta$ (rotation angle). The optimisation over $\theta$ is straightforward and yields $\theta=0$. \eq{dg2} then becomes
\begin{eqnarray}\label{mazza2o}
D_G(\gr\Sigma_{AB})&=&\inf_{m,\lambda} \left[\frac{1}{\sqrt{\frac{m^2 \left(a (b+\lambda  m)-c^2\right) \left(a (b \lambda +m)-d^2 \lambda \right)}{(b \lambda +m) (b+\lambda  m)}}}-\frac{4}{\sqrt{\frac{\left(a (b+\lambda  m)-c^2\right) \left(a (b \lambda +m)-d^2 \lambda \right)}{\lambda }}} \right. \nonumber \\ &+& \left.\frac{1}{\sqrt{\left(a b-c^2\right) \left(a b-d^2\right)}}\right]\,.
\end{eqnarray}

The remaining optimisations can be in principle solved analytically, although the resulting formulas are not very handy in general and there is no need to report them here explicitely. We prefer instead to specify our attention to the physically relevant subclass of two-mode squeezed thermal states, characterised by $d=\pm c$ in their covariance matrix $\gr\Sigma_{AB}^{\rm sts}$. For these states, the problem admits a simple solution, and the GGD turns out to be minimised at $$\lambda=1\,,\quad m= \frac{\sqrt{ab} \big(\sqrt{4 a b - 3 c^2} + \sqrt{a b }\big)}{3 a}\,.$$ The least disturbing  GPOVM for squeezed thermal states, according to the Hilbert-Schmidt distance, is thus a (noisy) heterodyne detection, a result which is analogous to what found in the case of quantum discord\cite{Giorda_10,AD_10}. The GGD of two-mode squeezed thermal states finally acquires the following compact expression
\begin{equation}
\label{ggdsts}
D_G(\gr\Sigma_{AB}^{\rm sts}) =
\frac{1}{a b -c^2} - \frac{9}{\left(\sqrt{4 a b - 3 c^2} + \sqrt{a b }\right)^2}\,.
\end{equation}
Notice that the case of pure two-mode squeezed states is recovered for $b=a, c=\sqrt{a^2-1}$.   A quite unexpected (yet nice) feature of \eq{ggdsts} is that $D_G$ of two-mode squeezed thermal states happens to be {\it symmetric} under party swap ($a \leftrightarrow b$), despite the generally nonsymmetric definition of the GGD.

We can test the accuracy  of the GGD in quantifying non-classical correlations of Gaussian states by means of a comparison with the conventional quantum discord ${\cal D}$ \cite{OZ}, whose definition and evaluation for two-mode Gaussian states can be found in detail in\cite{Giorda_10,AD_10}. Here, it suffices to recall that the (Gaussian) quantum discord ${\cal D}$ quantifies the minimal change in the mutual information of a bipartite (Gaussian) state after an optimal, least disturbing local (Gaussian) measurement is performed on subsystem $B$; notice that ${\cal D}$ can range up to infinity, while $D_G$ is normalised to one as already remarked.
 We run the analysis for squeezed thermal states, although the conclusions can be easily extended to more general two-mode Gaussian states.

\begin{figure}[tbh]
\centering
\subfigure[ ]{\includegraphics[width=10cm]{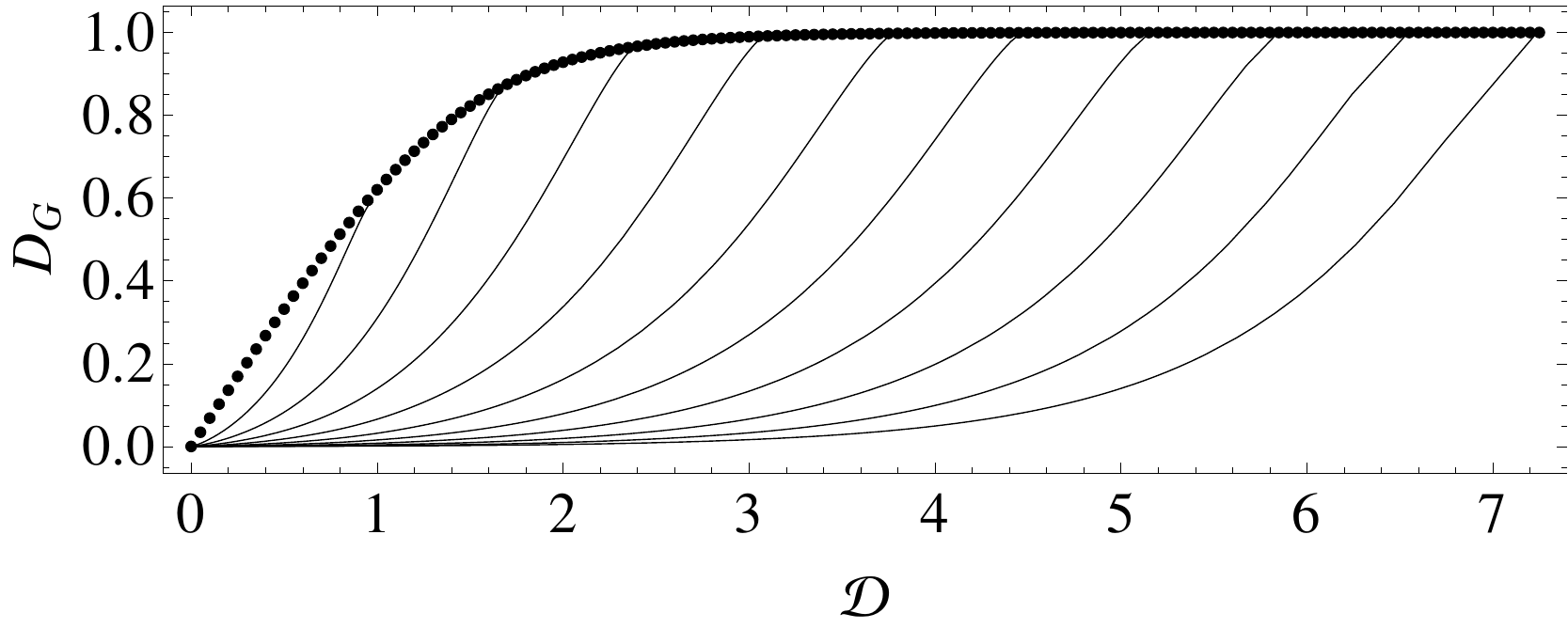}\label{figbona}} \\
\subfigure[ ]{\includegraphics[width=10cm]{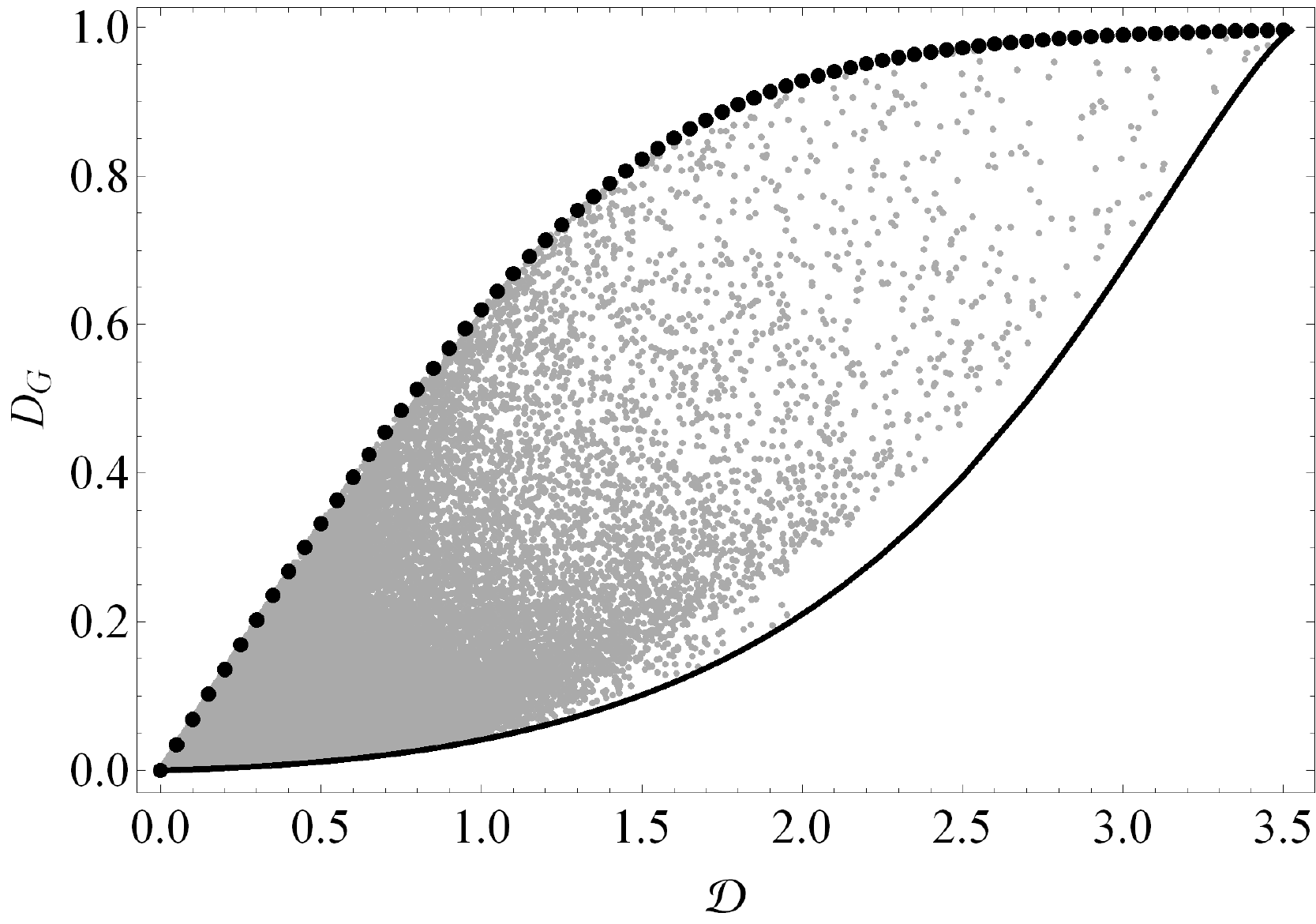}\label{figbonb}} \\
\caption{Upper (dotted black) and lower (solid black) bounds on the GDD $D_G$ at fixed quantum discord ${\cal D}$ for  two-mode squeezed thermal Gaussian states with (a) mean energy per mode bounded by $N=2^k$ with $k=1,2,\ldots,10$ (solid lines, from left to right) and (b) mean energy per mode bounded by $N=25$. In both plots, the lower boundaries (solid lines) correspond to states of the form \eq{cmlow}, while the upper boundary (dotted, independent of $N$) correspond to optimised states belonging to the class of \eq{cmup}, as detailed in the main text. Panel (b) includes a numerical exploration of $D_G$ versus ${\cal D}$ for $10^6$ randomly generated two-mode squeezed thermal states with mean energy per mode bounded by $N=25$ (gray points). All the quantities plotted are dimensionless.}
\label{figbon}
\end{figure}

A first observation that one can make, is that there is no {\it a priori} finite lower bound on $D_G$ for a fixed amount of quantum discord ${\cal D}$. This can be understood by recalling that in infinite-dimensional Hilbert spaces, infinitely entangled states (or, equivalently, states with diverging quantum discord\cite{AD_10}) constitute a (trace-norm) dense set. Our results show that the same holds for the Hilbert-Schmidt norm as well. This means, in particular, that in any neighbourhood of every product (classical) state $\chi_{AB}$ lies an arbitrarily strongly entangled (non-classical) state $\rho_{AB}$. The latter state $\rho_{AB}$ would thus have arbitrarily large quantum discord, and arbitrarily small geometric discord. We can provide an example family of two-mode squeezed thermal states that falls into this particular category. Consider the states with standard form covariances
\begin{equation}\label{cmlow}
b=1+\epsilon\,,\quad c=\sqrt{(a+1) \epsilon}\,, \quad \mbox{with $0\le\epsilon\le a-1$}\,.
\end{equation}
A plot of $D_G$ versus ${\cal D}$ is shown in Fig.~\ref{figbona} for the states of \eq{cmlow}, with $a=2^k$ and $k=1,2,\ldots,10$ (from left to right, solid lines). For these states,
${\cal D}(a,\epsilon) =
 \bigg[4 (a+1) \tanh ^{-1}\left(\frac{\epsilon }{2 a-\epsilon +2}\right)+2 \epsilon  (-a+\epsilon +1) \coth ^{-1}(a-\epsilon )-\epsilon  (\epsilon +2) \log \left(\frac{\epsilon }{\epsilon +2}\right)\bigg]/[2(2+\epsilon)]$ and $D_G(a,\epsilon)$ can be obtained from \eq{ggdsts}.
By setting $\epsilon=x(a-1)$ with a finite ratio $0<x<1$, one sees that, rigorously, $D_G \underset{a\rightarrow \infty}{\longrightarrow} 0$  for this class of Gaussian states, and ${\cal D} \underset{a\rightarrow \infty}{\longrightarrow} -x^{-1} \ln(1-x)$, which increases arbitrarily for $x$ approaching $1$, thus  demonstrating the claim. Let us remark that this feature is general,  only due to the  geometry of CV state spaces, and is not a peculiarity of our definition \eq{dgeommg} of $D_G$. Even allowing for any non-Gaussian measurement on $B$, or equivalently any non-Gaussian classical-quantum state in \eq{dgeom}, one could only further reduce the geometric discord, which would then still vanish for all those cases, like the two-mode family considered above, in which the GGD defined by \eq{dg2} vanishes  in the presence of an arbitrary degree of non-classical correlations measured by ${\cal D}$ (or by some entanglement measure).

A physically motivated way to address this issue is  by introducing a cap on the mean energy per mode of the Gaussian states under investigation\cite{eisertsimon}. This corresponds to imposing ${\rm tr} (\sig) \le 2N$ on the marginal covariance matrix of any single-mode Gaussian state, for some finite $N \ge 1$. The pathological behaviour in fact occurs in the unphysical limit of diverging mean energy. For squeezed thermal states in standard form, the finite energy constraint simply imposes $a,b \le N$. We have run an extensive numerical  comparison of $D_G$ versus ${\cal D}$ for $10^6$ randomly generated two-mode squeezed thermal states with bounded mean energy per mode (setting $N=25$), as shown in Fig.~\ref{figbonb}. In this case, as expected, a well definite region in the parameter space is filled by the random states, and an exact lower bound on the GGD can be identified at fixed quantum discord. Interestingly, it is saturated precisely by states of the form \eq{cmlow} with $a=N$.

We have also observed the presence of a general upper bound on $D_G$ versus ${\cal D}$ for two-mode squeezed thermal states.  It does not depend on the energy capping (i.e., it is independent of $N$) and thus stands as an upper limit on the GGD for any two-mode squeezed thermal state, even asymptotically infinitely entangled. The states saturating the upper bound can be found  within the two-parameter class characterised by \eq{sigsf} with
\begin{equation}\label{cmup}
a=b-\epsilon\,,\quad c=\sqrt{(b+1) (b-\epsilon-1)}\,, \quad \mbox{with $0\le\epsilon\le b-1$}\,.
\end{equation}
For these states, ${\cal D}(b, \epsilon)=\frac{1}{2} [(1-b) \ln (b-1)+(b+1) \ln (b+1)+\epsilon  \ln (\epsilon )-(\epsilon +2) \ln (\epsilon +2)]$ and $D_G(b,\epsilon)$ can be found again via \eq{ggdsts}. We are interested in a one-parameter subclass of this family, for which $D_G$ is maximised at a fixed value of ${\cal D}$. This optimisation can be performed numerically in two steps: first, identifying the contours of constant discord ${\cal D}$ in the space of parameters $b$ and $\epsilon$; then, maximising $D_G$ along each contour. The result is plotted as a dotted black curve in Fig.~\ref{figbon}. We notice that, for ${\cal D} \gg 0$, the maximal $D_G$ saturates quite rapidly to the maximal value of $1$, and the states sitting on the upper boundary converge simply to pure two-mode squeezed states, i.e., $\varepsilon=0$ in \eq{cmup} is optimal for highly correlated states. Again, this general result would still hold if the definition in \eq{dgeom} was employed.

Summarising, if no bound is imposed on the mean energy per mode, what we have found is that, essentially, for strongly non-classically correlated Gaussian states, the GGD is a rather insensitive measure of quantum correlations, as it can range arbitrarily between $0$ and $1$. On the other hand, for finite degrees of correlations (on entangled as well as separable states), respecting a physical cap on the mean energy, nontrivial upper and lower bounds exist on the GGD at fixed quantum discord, as demonstrated explicitly for the family of two-mode squeezed thermal states. We remark that exact upper and lower bounds on the geometric discord at fixed quantum discord have been recently demonstrated in the case of arbitrary states of two-qubit systems\cite{davide}.

\section{Geometric quadrature correlations} \label{secQ}

In this section we briefly discuss an alternative possible pathway to extend the notion of a geometric measure of non-classical correlations to bipartite Gaussian states. One might apply the notion of minimal Hilbert-Schmidt distance from the set of classical-quantum states, as in \eq{dgeom}, directly at the level of covariance matrices, as they provide a compact and complete characterisation of general Gaussian states up to local unitaries. We then define the ``geometric quadrature correlations'' (GQC) as
 \begin{equation}\label{gqc}
  Q_G(\gr\Sigma_{AB})= \inf_{\gr\Gamma_{AB} \in \Upsilon}  \|\gr\Sigma_{AB} -\gr\Gamma_{AB} \|_2^2\,.
  \end{equation}
where $\Upsilon$ denotes the set of all physical covariance matrices $\gr\Gamma_{AB}$ associated to classical-quantum Gaussian states with respect to the $A$ versus $B$ bipartition of modes. This set can be characterised following the results of Ref.~\cite{petz} (see also the Supplemental Material in\cite{AD_10}). The quantity $Q_G$ in \eq{gqc} is always nonnegative and manifestly invariant under local symplectic transformations, it vanishes on classical-quantum Gaussian states, and can range up to infinity.

For two-mode Gaussian states, $Q_G$ simply reduces to the minimum distance, in phase space, from completely uncorrelated tensor product states: $Q_G(\gr\Sigma_{AB})= \inf_{\sig_1, \sig_2}  \|\gr\Sigma_{AB} - (\sig_1 \oplus \sig_2) \|_2^2\,.$
As in the case of the GGD, we can evaluate this quantity explicitly assuming $\gr\Sigma_{AB}$ to be in standard form, \eq{sigsf}, and optimising the distance over arbitrary single-mode states $\sig_{1,2}$ of the form \eq{sigb}, specified by the sets of parameters $\{m_1, \lambda_1, \theta_1\}$ and $\{m_2, \lambda_2, \theta_2\}$, respectively. We then have
\begin{eqnarray}\label{q2m}
Q_G(\gr\Sigma_{AB}) &=& \inf_{\left\{
                                    m_{1,2},\lambda_{1,2},\theta_{1,2}\right\}} {\rm tr} \left[
                                  (\gr\Sigma_{AB} - \sig_1 \oplus \sig_2)^2\right] \,.
                                  \end{eqnarray}
                                  Explicitly,
                                  $Q_G(\gr\Sigma_{AB})=\inf_{\left\{
                                    m_{1,2},\lambda_{1,2},\theta_{1,2}\right\}}
                       \big[-2 b \lambda _2 \left(\lambda _2^2+1\right) \lambda _1^2 n_1+\left(\lambda _2^4+1\right) \lambda _1^2 n_1^2+\lambda _2^2 \left(2 \lambda _1^2 \left(a^2+b^2+c^2+d^2\right)-2 a \lambda _1 \left(\lambda _1^2+1\right) m_1+\left(\lambda _1^4+1\right) m_1^2\right)\big]/(\lambda _1^2 \lambda _2^2)$.

The expression is much more tractable than $D_G$ and there is no dependence on the angles $\theta_j$. The remaining optimisation over $m_j,\lambda_j$ can be easily solved by imposing vanishing partial derivatives with respect to the parameters, and one finally gets, for arbitrary two-mode Gaussian states in standard form, that $Q_G$ attains its global minimum for $\lambda_1=\lambda_2=1$, $m_1=a$, $m_2=b$. In other words, the closest classical state is just the tensor product of the two marginals, $\gr\Gamma_{AB} = \gr\alpha  \oplus \gr\beta$. The GQC thus acquires a very simple form:
\begin{eqnarray}\label{gsimple}
Q_G(\gr\Sigma_{AB}) &=& 2(c^2+d^2) \\
&=& \frac{2(\det\gr\alpha+\det^2\gr\gamma-\det\gr\Sigma_{AB})}{\sqrt{\det\gr\alpha \det\gr\beta}}\,, \nonumber
\end{eqnarray}
where in the second line we have explicitly rewritten the solution in terms of the invariants of the covariance matrix [\eq{sig}],  in order to provide a general formula not relying on the standard form.

The GQC is then a symmetric measure that obviously quantifies {\it total} correlations of two-mode Gaussian states, as intuitively clear from the definition \eq{q2m}. Its standard form expression is particularly transparent, as it just coincides with the sum of the squared off-diagonal quadrature correlations in the state characterised by $\gr\Sigma_{AB}$, which essentially account for all the possible correlations of Gaussian states (whose higher order correlations in the canonical phase-space operators are entirely specified by the ones between second moments). We notice a similarity between this formula and its counterpart applied at the level of density matrices for finite-dimensional systems, where {\it quantum} correlations can be quantified by the so-called ``minimum entanglement potential'' (or negativity of quantumness), which amounts to the sum of the modulus square of all the off-diagonal coherences in the density matrix of a bipartite system, minimised over all possible local bases\cite{genio}.

 We clarify with an example how GQC in the Gaussian setting quantifies the amount of total (i.e. both classical and quantum) correlations. Let us consider the family of two-mode Gaussian states with standard form covariances given by
\begin{equation}
\label{extot}
b=a\,,\quad c=a-1\,,\quad d=0\,.
\end{equation}
For these states, both the quantum discord\cite{AD_10} ${\cal D}$ and the GGD $D_G$ [\eq{dg2}] are very small for any $a \ge 1$, never exceeding $1/40$ and vanishing exactly in the limit $a \rightarrow \infty$. On the contrary, the GQC quantity $Q_G$ scales as $2(a-1)^2$, diverging for large $a$. The measure $Q_G$ is then closer in spirit to the ``bit quadrature correlations'' studied in\cite{carles} for two-mode CV states.

\section{Conclusions} \label{secC}

In this paper we addressed the quantification of general non-classical correlations in Gaussian states of CV systems from a geometric perspective. Generalising the known finite-dimensional definitions\cite{dakic,luofu}, we proposed a Gaussian version of the geometric measure of discord, defined as the minimum distance between a bipartite Gaussian state and the closest classical-quantum Gaussian state obtained after a local Gaussian measurement on one subsystem only. We calculated the measure for two-mode Gaussian states, focusing in particular on squeezed thermal states. We then compared the Gaussian geometric discord with the conventional Gaussian quantum discord\cite{OZ,Giorda_10,AD_10},  and identified upper and lower bounds on one quantity as a function of the other,  upon constraining  the mean energy per mode. In absence of such a provision, the geometric discord can vanish for arbitrarily strongly non-classical (entangled) Gaussian states, as a consequence of the geometry of CV state spaces\cite{eisertsimon}.

 We finally analysed a different Gaussian counterpart to the geometric discord, where the distance from the set of classical-quantum states is calculated directly on covariance matrices; this operation results in a simple quantifier that captures the total quadrature correlations rather than the genuine non-classical correlations of Gaussian states.

Contrarily to the two-qubit case, where the geometric quantum discord is easier to calculate than the conventional discord\cite{dakic,luofu,davide}, we find no significant advantage (beyond the fact of managing polynomial rather than logarithmic functions) in choosing a geometric perspective rather than an entropic one, for the quantification of Gaussian non-classical correlations. It would be interesting to investigate alternative distance-based measures of non-classical correlations for Gaussian states, based on different norms, e.g. in terms of phase-space distinguishability measures.

We finally mention that, following the analysis of\cite{dakic,bogna}, quantum discord has been very recently witnessed in qubit setups implemented with nuclear magnetic resonance\cite{laflammetal}. Inspired by that, an interesting future task could be trying to devise proper quantitative witnesses of Gaussian (geometric) quantum correlations, accessible experimentally in all-optical implementations or hybrid setups interfacing light modes and ultracold atomic ensembles.

\noindent {\bf Acknowledgments.} {We acknowledge the University of Nottingham for financial support through an Early Career Research and Knowledge Transfer Award.}

\end{document}